\documentclass[a4paper,11pt]{article}

\usepackage{graphicx}
\usepackage{multirow}
\usepackage{hhline}
\usepackage{times}
\usepackage{lipsum}
\usepackage{wrapfig}

\let\OLDthebibliography\thebibliography
\renewcommand\thebibliography[1]{
  \OLDthebibliography{#1}
  \setlength{\parskip}{0pt}
  \setlength{\itemsep}{0pt plus 0.3ex}
}

\begin{document}

{\hskip -0.6cm
\begin{minipage}[h]{16cm}
\begin{center}
\textbf{IONOSPHERIC D-REGION TEMPERATURE RELAXATION \\ AND ITS INFLUENCES ON RADIO SIGNAL PROPAGATION \\AFTER SOLAR X-FLARES OCCURRENCE}
\vspace{\baselineskip}

\textit{\textbf{Jovan~BAJ\v{C}ETI\'{C}$^{a*}$,
        Aleksandra~NINA$^{b}$,
        Vladimir~M.~\v{C}ADE\v{Z}$^{c}$
        and~Branislav~M.~TODOROVI\'{C}$^{d}$}}
        \vspace{\baselineskip}

$^{a}$ Department of Telecommunications and Information Science, University of Defence, Military Academy, Serbia\\
$^{b}$ Institute of Physics, University of Belgrade, Serbia\\
$^{c}$ Astronomical Observatory of Belgrade, Serbia\\
$^{d}$ RT-RK Institute for Computer Based Systems, Serbia\\

\vspace{\baselineskip}
\end{center}
\end{minipage}


}


{
\hoffset -3.6cm
\hskip 1.35cm
\begin{minipage}[h]{12cm}
{\textbf{Abstract.} \textit{In this paper our attention is focused on relations between radio signal propagation characteristics and temperature changes in D-region after solar X-flare occurrence. We present temperature dependencies of electron plasma frequency, the parameter that describes medium conditions for propagation of an electromagnetic wave, and the refractive index which describes how this wave propagates. As an example for quantitative calculations based on obtained theoretical equations we choose the reaction of the D-region to the solar X-flare occurred on May 5$^{ \rm {th}}$, 2010. The ionospheric modelling is based on the experimental data obtained by low ionosphere observations using very low frequency radio signal.}

}
 \end{minipage}
}


\let\thefootnote\relax\footnote{{$^*$}Corresponding author; e-mail: jovan.bajcetic@va.mod.gov.rs}

%

\section{Introduction}

As a part of the atmosphere, the ionosphere and particularly it's  lowest region, the D-region (60 km - 90 km altitude), is under permanent influences coming from the outer space, other regions of atmosphere and lithosphere. For this reason, the characteristics of local plasma are time dependent as reported in numerous studies related to the electron density and other plasma parameters changes induced by solar X-flares \cite{nin11,sch13}, solar eclipse \cite{sin11b}, and lightning \cite{ina88}, or induced harmonic and quasi-harmonic hydrodynamic motions (including soliton formation and vortices) \cite{jil13,nin13a}.
These perturbations cause changes in thermal properties of the considered medium followed by temperature variations in the ionosphere \cite{ina91,sha11}. In the case of ionospheric heating induced by astrophysical events, more intensive changes of electrons and ions temperature in the ionosphere happen at the higher altitudes, e.g. in the F-region. Because of that, the published studies related to these plasma parameter variations usually present the conditions in this region \cite{smi06}. We have not found adequate reports for the D-region temperature time evolution in literature which was the motivation to expand investigation of temperature variation to lower altitudes.

In addition to pure scientific significance, and possible applications in prediction of elementary disasters, ionospheric investigation can be of practical use in the field of radio telecommunications \cite{kha14}. Namely, radio wave propagation over long distances is enabled by multiple reflections of electromagnetic (EM) waves inside the Earth - ionosphere waveguide and is dependent of ionospheric properties which can be changed by X-flares, Lyman-$\alpha$ radiation, lightnings, geomagnetic storms and many other effects.
The daytime low ionospheric variations have the greatest affect on radio wave propagation at low (30 kHz - 300 kHz), very low (3 kHz - 30 kHz) and ultra low (0.3 kHz - 3 kHz) frequency bands \cite{ina07,sul14}. Prediction of such variations is important for the variety of reasons: remote sensing detection of narrow bipolar events in clouds \cite{ush14}, monitoring of acoustic and gravity waves in the atmosphere \cite{nin13a}, and monitoring specific perturbations in the ionosphere that precede seismic activity \cite{hay10} and tropical cyclones \cite{pri07}. The parameters describing EM waves and characteristics of medium important for radio wave propagation are: the refractivity index and the electron plasma frequency, respectively. In this study we deal with relations of these parameters with temperature time evolutions within the D-region.

The analysis of ionospheric variations is rather difficult because of space and time dependent variable influences arising from numerous geophysical, astrophysical and artificial events. For this reason, experimental observations play a very important role in more realistic modelling. Method selection depends on the considered altitude. Thus, in the case of low ionosphere, there are three suitable techniques based on very low frequency (VLF) radio waves, rocket and radar measurements \cite{str10,nin12a,cha14}. The first of them, which is used in this work is based on continually emitted and recorded signals by numerous, worldwide distributed, transmitters and receivers, respectively. In addition to detections of periodic and global low ionospheric perturbations, such an experimental setup enables registrations of sudden and unpredictable local plasma variations like those induced, for example, by lightnings or earthquakes.

In this paper our attention is focused on two topics. First, we give a study of temperature relaxation after large perturbations. Here, we introduce a new theoretical procedure to calculate temperature time evolution at fixed altitudes based on experimental VLF data and LWPC (Long-Wave Propagation Capability) numerical model for the VLF signal propagation simulation \cite{fer98} used for the D-region plasma modelling. This procedure follows the research presented in \cite{nin14} and reference therein and, as we already said, it extends the temperature time evolution research to D-region heights. In the second part we focus on the propagation of radio waves in considered conditions and we show how the electron plasma frequency (describing the medium characteristics important for EM waves propagation) and refractive index (representing the main aspect of EM wave propagation) are related to the electron temperature at different altitudes.

The obtained theoretical equations are applied to a particular case of the D-region perturbation induced by solar X-flare occurred on May 5$^{ \rm {th}}$, 2010.

\section{Experimental setup and observed data}

In this study, we based our analysis on low ionospheric monitoring using the 23.4 kHz VLF signal emitted by the DHO transmitter located in Rhauderfehn (Germany) and received at Institute of Physics in Belgrade (Serbia). This transmitter was chosen in many studies because it provides the best quality suitable signal frequency for the location of the receiver, and has a relatively short signal propagation path.

\begin{wrapfigure}[24]{r}[0pt]{0.45\textwidth}
\vspace{-12pt}
\begin{center}
\vspace{-10pt}
\includegraphics[width=0.45\textwidth]{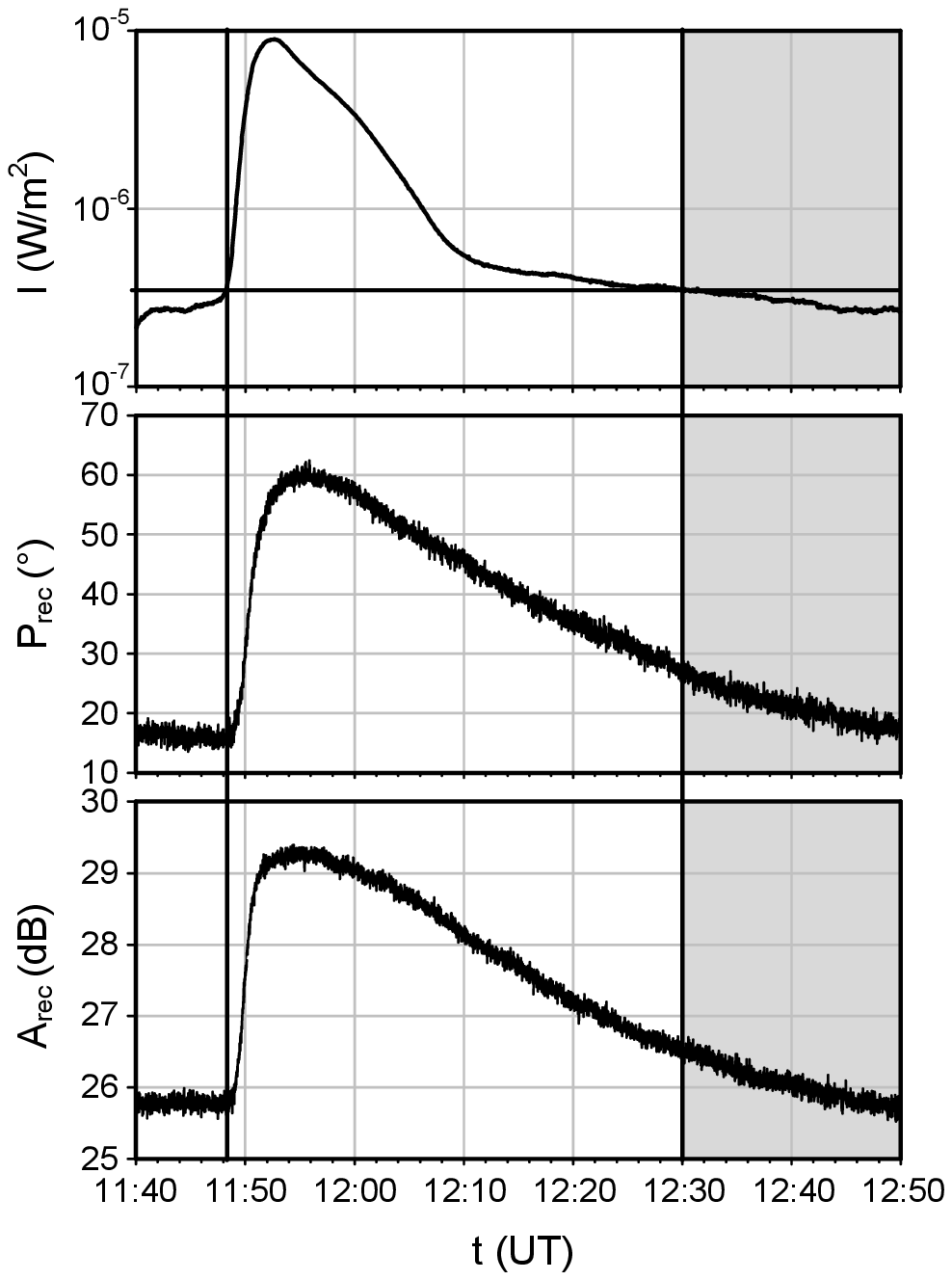}
\end{center}
\vspace{-15pt}
\caption{\small{Increase of X-radiation registered by the GOES-14 satellite in wavelengths domain 0.1 nm - 0.8 nm (upper panel) and reaction of phase (middle panel) and amplitude (bottom panel) of the VLF signal emitted by the DHO transmitter located in Germany and received by the AWESOME receiver in Serbia. The shaded domain represents the relaxation period analyzed in this study.}}
\vspace{-10pt}
\label{fig_rec}
\end{wrapfigure}

The final theoretical results of presented study and required numerical procedure for modelling the D-region plasma are applied in a particular case of perturbation induced by the solar X-flare occurred on May 5$^{ \rm {th}}$, 2010 with photon flux $I$ registered by the GOES-14 satellite of National Oceanic and Atmospheric Administration (NOAA), USA (Fig. \ref{fig_rec}, upper panel) at wavelengths range 0.1 nm - 0.8 nm. The ionospheric perturbations were detected as amplitude $\Delta A _{\rm {rec}}$ and phase $\Delta P _{\rm {rec}}$ variations of the considered VLF signal recorded by the AWESOME (Atmospheric Weather Electromagnetic System for Observation Modelling and Education) VLF receiver (bottom and middle panels, respectively).

In Fig. 1, the time evolutions are shown for the whole perturbation period. However, the presented study is related to the relaxation period of the described X-radiation influence. This period is indicated by the shaded domain whose start is determined by the moment when the intensity $I$, after passing thorough maximum, reaches the same value as it was at the time when the signal characteristics started to response to the considered event. Here, we assume that the smaller intensity of X-radiation does not have a significant role in the D-region ionization which is important in explanation of the electron density dynamics (see Eq. (\ref{eq:5}) and its explanation).

\section{D-region modelling}
\label{modelling}


As we said in Introduction, in the first part of this article, we analyze temperature relaxation in the D-region after large perturbations which induce a relatively long time period of plasma relaxation. Generally, there are different models for low ionospheric parameter determinations. In this paper, we expand investigations of the D-region plasma based on its real time monitoring that allows us to use data related to the considered time period and location.

In the second part, we compare the obtained temperature changes with modelling parameters of the D-region that are important for propagation of telecommunication signals.

\subsection{Modelling temperature time evolutions}

Our theoretical procedure is based on calculations of temperature time evolution from the effective recombination coefficient $\alpha_{eff}$. This coefficient varies in time with significant differences between values in the daytime and nighttime periods. Here, we considered daytime conditions and expressed $\alpha_{eff}$ through recombination coefficients $\alpha_{i}$ related to ion whose role is important in the D-region electron loss processes, and electron $N_e$ and ions $N_i$ densities \cite{mit77}:

\leavevmode
\begin{equation}
\label{eq::alfaeffi}
\alpha_{eff}=\frac{1}{N_e}\sum _i \alpha _i N_i.
\end{equation}

The results of investigations show that the temperature dependencies of the recombination coefficients can be given as $\alpha_i=C_i \cdot (T_e/300)^{D_i}$ where $T_e$ is electron temperature and coefficients $C_i$ and $D_i$ depend on ion species \cite{rod98}. The relevant values for N${_2}{^+}$, O${_2}{^+}$, NO${_2}{^+}$ and cluster ions H$^+$(H$_2$O)$_n$, given in \cite{rod98} and reference therein, are shown in Table \ref{table1}.

\begin{wraptable}[10]{r}[0pt]{0.4\textwidth}
\vspace{-25pt}
\caption{\small{Coefficients $C_i$ and $D_i$ for ions which has important role in recombination process within the D-region.}}
\begin{center}
\leavevmode
\vspace{-20pt}
\begin{tabular}{|c||c|c|}
\hline
   Ion & $C _ i$ (10$^{-13}$) & $D _ i$ \\
   \hline
   \hline
   {N$_2$}$^+$ & 1.8 & -0.39 \\
   \hline
   {O$_2$}$^+$ & 1.6 & -0.55 \\
   \hline
   {NO}$^+$ & 4.5 & -0.83 \\
   \hline
   H$^+$(H$_2$O)$_n$ & 5+20n & -0.5 \\
   \hline
\end{tabular}
\end{center}
\label{table1}
\end{wraptable}

Contributions of particular ion species in recombination processes depend on the altitude. At higher D-region altitudes the dominant role have {O$_2$}$^+$ ions while, below 80 km, a significant increase of influence of cluster ions is found \cite{rod98}. Because of that, in the second case, it can be taken that the coefficient $\alpha_{eff}$ is proportional to $(T_e/300)^{-0.5}$. In our model, we express Eq. (\ref{eq::alfaeffi}) as

\leavevmode
\begin{equation}
\label{eq::alfaeffi1} \alpha_{eff}=\alpha _{Claster}(1+\sum _{i'}
r_{\alpha _i}^{Claster} r_{N_i}^{Claster}),
\end{equation}
where $r_{\alpha _i}^{Claster}$ and $r_{N_i}^{Claster}$ represent the ratios of recombination coefficients and densities of ions $i$ and claster, respectively. Therefore, in addition to taking into account the dominant processes in the relaxation which are studied in \cite{sch13,rod98,rod07} we also considered variable altitude contributions of less significant processes. At the end of relaxation, time variations of densities are very small. Also, the $r_{\alpha_{O_2}}^{Claster}$ are changed by less than 1\% in temperature range 190 K - 230 K (here, we use $\alpha_{Claster} = 7.5\cdot10^{-12}\sqrt{300/T_e}$, see \cite{rod98}). Under these conditions, we can consider the expression in parentheses in the previous equation to be approximately constant at some altitudes and write Eq. \ref{eq::alfaeffi1} as:

\leavevmode
\begin{equation}
\label{eq::alfaodT} \alpha_{eff}=C(T_e/300)^{-0.5},
\end{equation}
where $C$ can be calculated from the effective recombination coefficient $\alpha_{eff}^0$ and electron temperature $T{_e}^0$ under unperturbed conditions:

\leavevmode
\begin{equation}
\label{eq::C} C=\alpha_{eff}^0(T_e^0/300)^{0.5}.
\end{equation}

Finally, from Eq. \ref{eq::alfaodT} we obtain the expression for the electron temperature time evolution:

\leavevmode
\begin{equation}
\label{eq::Todt} T_e=300(\alpha_{eff}/C)^{-0.5}.
\end{equation}

As we can see, the presented model to determine the temperature time evolution at a given location required two initial space-time dependent quantities: the effective recombination coefficient $\alpha_{eff}^0$ and electron temperature $T{_e}^0$. These parameters can be calculated by different procedures. In this paper, we present our new way to determine $\alpha_{eff}$ determination, while $T{_e}^0$ is obtained from the IRI model \cite{bil92}.

%

\subsubsection{Modelling the relaxation effective recombination coefficient}

The effective recombination coefficient, and consequently its values $\alpha_{eff}^0$ in the unperturbed D-region ($dN_e/dt=0$) can be calculated from the equation for electron density dynamic \cite{mce78}:

\leavevmode
\begin{equation}
\label{eq:5}
    {
    \frac{dN_e(\vec{r},t)}{dt}= {\cal G}_0(\vec{r},t)-\alpha_{eff}(\vec{r},t)N_e^{2}(\vec{r},t).
    }
\end{equation}

Here, ${\cal G}_ {\rm 0}(\vec{r},t)$ is the electron gain rate in the unperturbed D-region at location $\vec{r}$ and at time $t$.
  If contributions of X-radiation and other sudden influences are not significant in ionization process, the electron gain rate can be used as a very slow time dependent quantity at the end of the D-region perturbation process. The procedure for its calculation is presented in \cite{nin14}. Also, we take that the density of negative ions is neglected with respect to electron density which allows us to consider recombination processes as the dominant electron loss processes \cite{mce78}.

The calculations of electron density are done by comparison of the recorded amplitude $\Delta A _{\rm {rec}}$ and phase $\Delta P _{\rm{rec}}$ changes with corresponding values obtained by the LWPC (Long-Wave Propagation Capability) numerical model for the VLF signal propagation simulation \cite{fer98} as explained in \cite{gru08}. The procedure is based on finding the combination of input parameters, the signal reflection height $H'$ (in km) and sharpness $\beta$ (in km$^{-1}$), that gives the best agreement of the recorded with modeled signal characteristics. The electron density $N_e(h,t)$ (in m$^{-3}$) at fixed altitude $h$ (in km) is calculated from these parameters by applying Wait's model of the ionosphere considering the vertical electron density gradient \cite{wai64}:

\begin{equation}
\label{eq::Ne}
N_e(h,t) = 1.43\cdot10^{13}e^{-\beta(t)H'(t)}e^{(\beta(t)-0.15)h},
\end{equation}
which is the standard model applied in numerous studies.

The data of the electron density time distribution in the considered time period and electron gain rate in the unperturbed D-region allow us to calculate the effective recombination coefficient $\alpha_{eff}$ from a relation derived from Eq.(\ref{eq:5}) for a horizontally uniform low ionosphere:

  \begin{equation}
    \label{eq:alfa}
    {
    \alpha_{eff}(h,t)= \frac{{\cal
    G}_0(h,t)-\frac{dN_e(h,t)}{dt}}{N_e^{2}(h,t)}.
    }
  \end{equation}

\subsection{Signal propagation and electron temperature}

In the second part of this study, our attention is directed to relations between the obtained temperature and parameters that describe radio signal propagation in the considered ionospheric part: the electron plasma frequency $f_{\rm {0}}$ and refractive index $n$ that can be obtained from  approximative relations \cite{bud88,dav89}:

\begin{equation}
\label{eq::f0}
f_{\rm {0}} \approx 8.98 \sqrt{N_e} \quad \rm{[Hz]}
\end{equation}

and

\begin{equation}
\label{eq::n}
n(h,t) = \sqrt{1-f_0^2/f^2},
\end{equation}
where $f$ is the considered radio signal frequency, and $N_e$ is given in m$^{-3}$. Here, we assume that there is no effect of the Earth's magnetic field. In reality it can significantly affect the propagation of radio waves causing signal distortion, but this is more important at higher ionospheric regions and for higher latitudes. As we will see from the obtained values for $f_{\rm {0}}$, significant changes in propagation are expected for the LF (30 - 300 kHz) and partly MF (300 - 500 kHz) radio signals which was the reason to consider these frequency domains in our analysis.

\section{Results and discussion}

\begin{wrapfigure}[13]{r}{0.5\textwidth}
\vspace{-70pt}
\begin{center}
\includegraphics[width=0.5\textwidth]{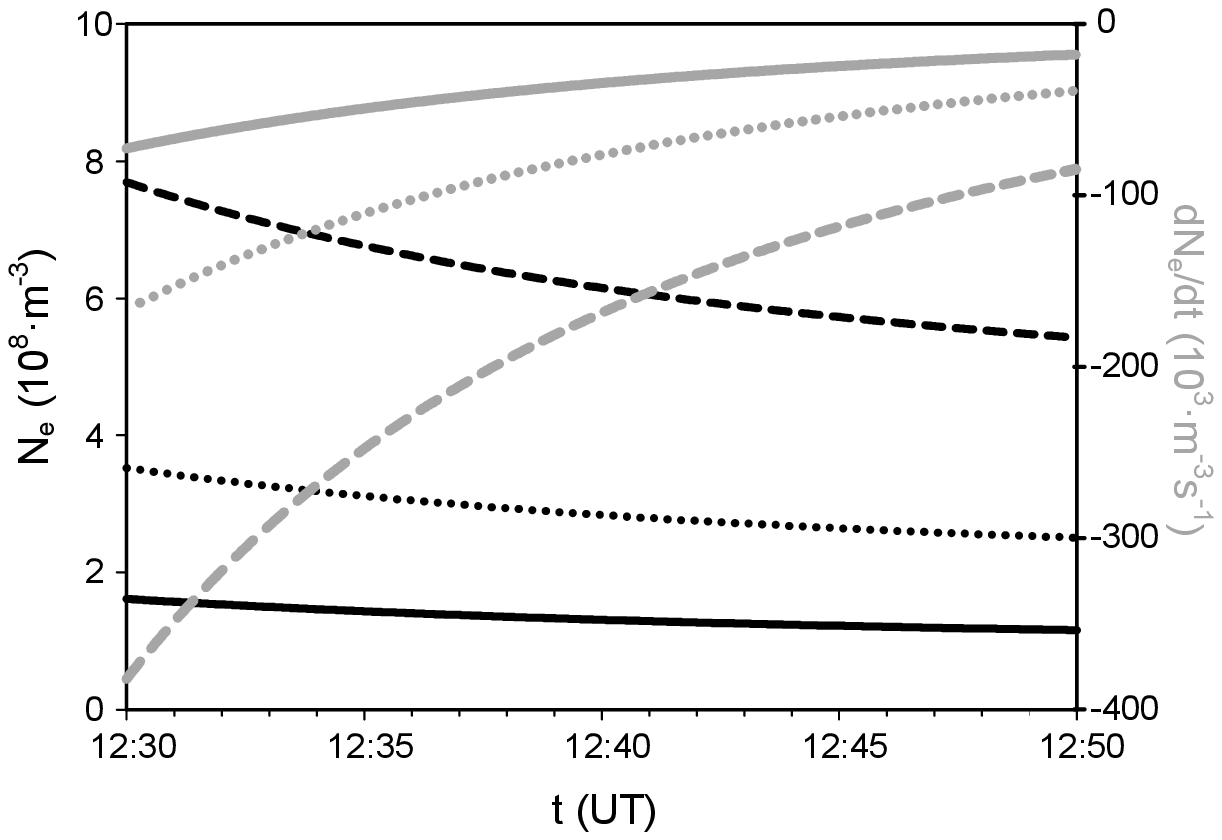}
\end{center}
\vspace{-20pt}
  \caption{\small{Time evolutions of electron density (black lines) and its time derivative (gray lines) at altitudes 70 km (solid line), 75 km (dot line) and 80 km (dash line) during relaxation.}}
  \vspace{-10pt}
  \label{Fig2_Nodt}
\end{wrapfigure}

The presented model gives an analytical expression for calculations of the temperature time evolution at fixed location after an ionospheric perturbation, using the time and space dependent initial data: the effective recombination coefficient $\alpha_{eff}^{\rm {0}}(h,t)$ and electron temperature $T_e^{\rm {0}}(h,t)$ under unperturbed conditions. Here, we first present the results related to determinations of initial data, and then time evolutions of electron temperature at fixed altitudes and its relations to electron plasma frequency $f_{\rm {0}}(h,t)$ and refractive index $n(h,t)$.
\begin{sloppypar}
\begin{wraptable}[10]{r}[0pt]{0.7\textwidth}
\vspace{-17pt}
\caption{\small{Values of electron gain rate ${\cal G}_0$, ambient temperatures in unperturbed ionosphere $T_e^{\rm {0}}$, effective recombination coefficient $\alpha_{eff}^{\rm {0}}$ and coefficient $C$, related to altitude.}}
\begin{center}
\leavevmode
\begin{tabular}{|c||c|c|c|c|}
\hline
   $Altitude$  & ${\cal G}_0$ \cite{nin14} & $T_e^{\rm {0}}$ \cite{bil92} & $\alpha_{eff}^{\rm {0}}$  & $C$  \\
   (km) &(m$^{-3}$s$^{-1}$) & (K) & (10$^{-12}$m$^{3}$s$^{-1}$) & (10$^{-12}$) \\
   \hline
   \hline
   70 & 41841 & 219.1 & 4.55 & 3.89\\
   \hline
   75 & 95000 & 205.4 & 2.13 & 1.76\\
   \hline
   80 & 215697 & 192.4 & 1.02 & 0.82\\
   \hline

\end{tabular}
\end{center}
\label{table2}
\end{wraptable}
\end{sloppypar}

\subsection{Determination of initial data}


Determination of the effective recombination coefficient $\alpha_{eff}(h,t)$ requires data analysis of electron density $N_e(h,t)$ and its time derivative which can be obtained from observed data by the procedure explained in \cite{gru08}. These dependencies, obtained from Eq. (\ref{eq::Ne}) in calculations presented in \cite{nin14}, are shown in Fig. \ref{Fig2_Nodt} for altitudes 70 km, 75 km and 80 km.
The given analysis is based on observation data obtained by the VLF monitoring of the D-region and modelling by the LWPC numerical model. The presented results show small variations of relevant values within the considered time period, which is important for the assumption of the time constant coefficient $C$ whose values computed from Eq. (\ref{eq::C}) are given in Table \ref{table2} for the considered altitudes. The values for $\alpha_{eff}^{\rm {0}}={\cal G}_0(h,t)/N_e^{2}(h,t)$ (in Eq. \ref{eq:5} for $dN_e/dt=0$), $T_e^{\rm {0}}$ and ${\cal G}_0(h,t)$ are taken from literature and quoted in the table.

\begin{wrapfigure}[25]{r}{0.5\textwidth}
\vspace{-55pt}
\begin{center}
\vspace{-20pt}
\includegraphics[width=0.5\textwidth]{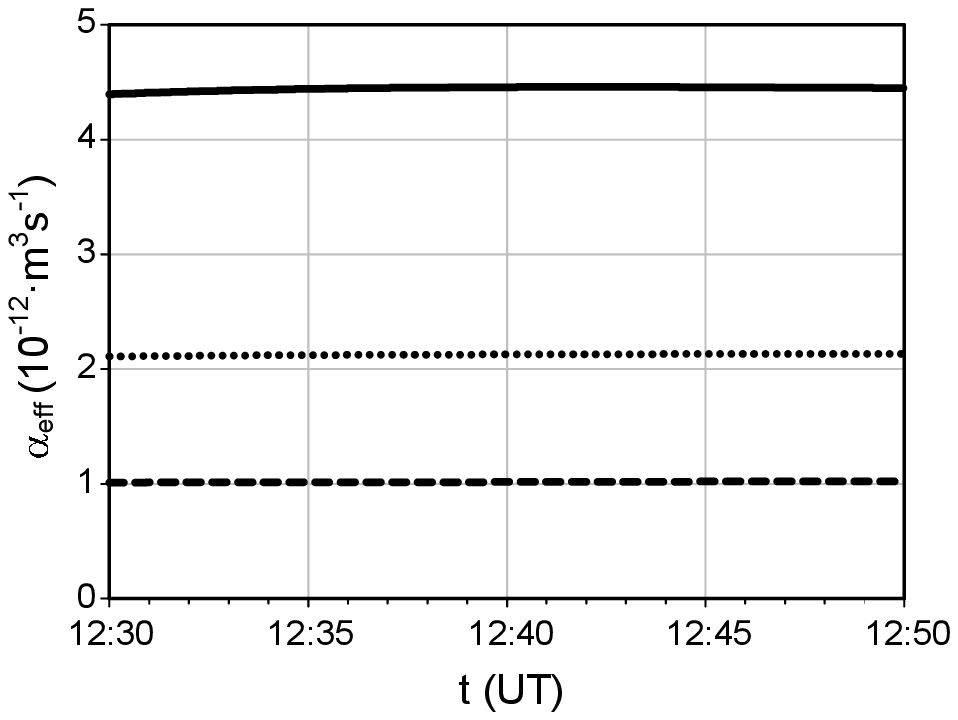}
\end{center}
\vspace{-15pt}
\caption{\small{Time evolutions of effective recombination coefficient at altitudes 70 km (solid line), 75 km (dot line) and 80 km (dash line) during relaxation.}}
\vspace{-10pt}
\label{Fig3_alfaodt}
\vspace{10pt}
\begin{center}
\vspace{-0pt}
\includegraphics[width=0.5\textwidth]{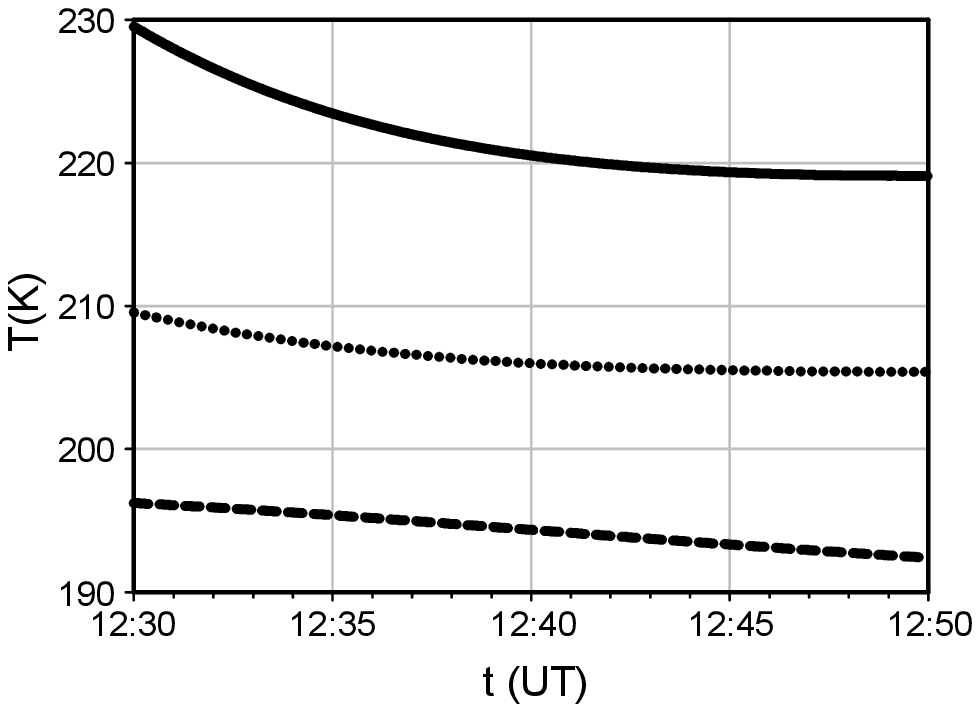}
\end{center}
\vspace{-20pt}
\caption{\small{Electron temperature time evolutions at altitudes 70 km (solid line), 75 km (dot line) and 80 km (dash line) during relaxation.}}
\vspace{-10pt}
\label{Fig4_Todt}
\end{wrapfigure}

Introducing the obtained data into Eq. (\ref{eq:alfa}) we get  time dependencies of the effective recombination coefficient at altitudes presented in Fig. \ref{Fig3_alfaodt}. Here, we see slow increases of this coefficient going to the unperturbed domain which are more visible at lower altitudes. Such a tendency is also found in \cite{nin12b}. The saturated values at the end of the relaxation time period are in a good agreement with those presented in literature (\cite{mit77}, \cite{har05} and references within).

%


\subsection{Temperature time evolutions}

Introduction of the obtained initial data into Eq. (\ref{eq::alfaodT}) gives the time evolution of temperature presented in Fig. \ref{Fig4_Todt} for altitudes: 70 km, 75 km and 80 km.  Those values indicate a smaller temperature variation at bigger heights. Also, the average rate of change and instantaneous change for temperature (presented in Table \ref{table3}) during the considered time period (20 min) show small variations of relevant values which allow us to use the presented method under the assumed conditions when $C$ is considered constant in time, which is explained in the Section \ref{modelling}.

%

\subsection{Temperature dependencies of quantities important in telecommunications}

\begin{wrapfigure}[18]{r}{0.5\textwidth}
\vspace{-20pt}
\begin{center}
\includegraphics[width=0.5\textwidth]{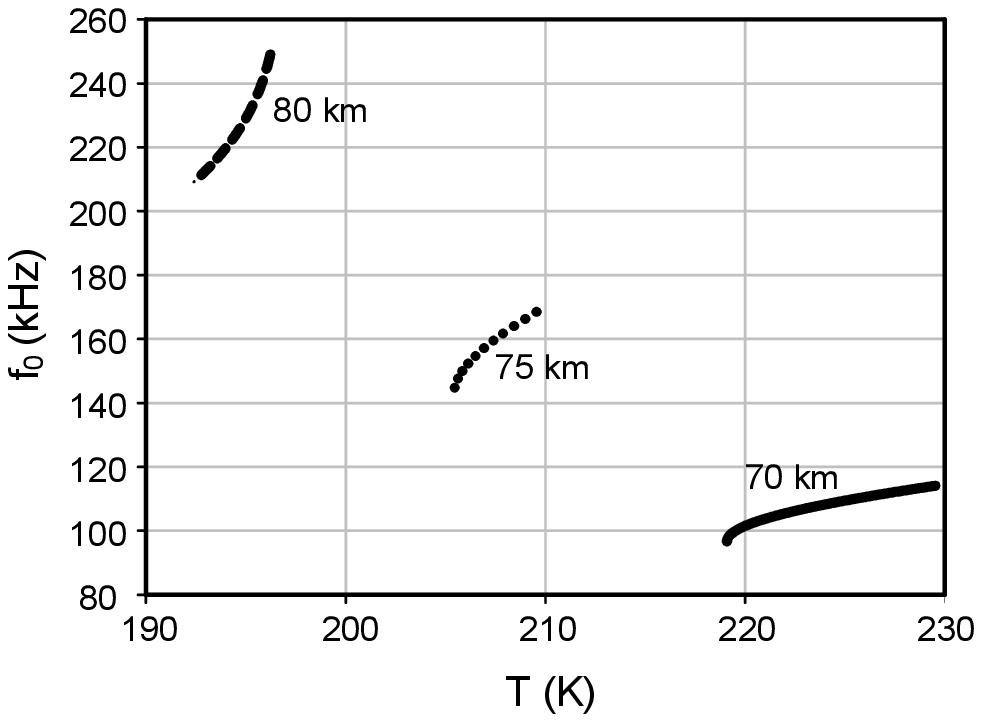}
\end{center}
\vspace{-20pt}
\caption{\small{Dependencies of electron plasma frequencies from electron temperature at altitudes 70 km (solid line), 75 km (dot line) and 80 km (dash line) during relaxation. The shown values represent the EM wave carrier frequencies for considered altitudes when total signal absorption (blackout) occurs. }}
\vspace{-10pt}
\label{Fig5_f0odT}
\end{wrapfigure}

Eq. \ref{eq::f0} shows that the electron plasma frequency, the characteristic parameter of the medium characterizing propagation of radio signals, becomes more strongly temperature dependent at higher altitudes as plotted in Fig. \ref{Fig5_f0odT}. Here, we can see different variations of the considered parameters at fixed altitudes which indicate their different relaxation behaviors after a solar X-flare. Our model gives larger relative variations of the electron plasma frequency (around 15.27\%, 15.64\% and 16.01\% at 70 km, 75 km and 80 km, respectively) with respect to temperature changes (-4.55\%, -1.99\% and -1.95\% at 70 km, 75 km and 80 km, respectively). From the average
rate of change and instantaneous change of $f_{\rm {0}}$,
presented in Table 3, we can see that these variations are not large and that the absolute values of the last quantity are larger then those in the case of temperature.

\begin{sloppypar}
\begin{wraptable}[12]{r}[0cm]{1\textwidth}
\vspace{-20pt}
\caption{\small{Average rates of temperature (${\Delta T / \Delta t}$) and electron plasma frequency (${\Delta f_{\rm {0}} / \Delta  t}$) changes, instantaneous relative change of temperature (${(\Delta T / \Delta t)/T}$ ) and electron plasma frequency ($({\Delta f_{\rm {0}} / \Delta  t})/f_{\rm {0}}$), related to altitude.}}
\vspace{0pt}
\begin{center}
  \begin{tabular}{|c||c|c|c|c|c|c|}
    \hline
    &  \multicolumn{2}{c}{$Average$ $change$ $rate$} & \multicolumn{2}{|c|}{$Instantaneous$ $relative$ $change$} \\
    \hhline{~------}

    {$Altitude$} &  ${\Delta T / \Delta t}$ & ${\Delta f_{\rm {0}} / \Delta  t}$ & ${(\Delta T / \Delta t)/T}$ & $({\Delta f_{\rm {0}} / \Delta  t})/f_{\rm {0}}$ \\
    \hhline{~~~~~~~}
    (km) &  (K/min) & (kHz/min) & (\%/min) & (\%/min) \\
    \hline
    \hline
    70 &  -0.522 & -0.871 & -0.227 & -0.764\\
    \hline
    75 &  -0.208 & -1.318 & -0.099 & -0.782 \\
    \hline
    80 &  -0.191 & -1.994 & -0.097 & -0.801 \\
    \hline
  \end{tabular}
\end{center}
\label{table3}
\end{wraptable}
\end{sloppypar}

From Fig. \ref{Fig5_f0odT} we see their opposite tendency with altitude: the  electron plasma frequency variations increase with altitude, while the corresponding temperature variations decrease with altitude. This results can be explained by a more intensive increase in electron density at higher layers during the radiation influence in the first case, and by larger variations of effective recombination coefficient at lower layers (see Fig.~4) in the second case.

%

The obtained values for $f_{\rm {0}}$ are further used in Eq. \ref{eq::n} for calculations of the medium refractive index change, as a function of temperature and communication signal carrier frequency for three referent altitudes in the ionospheric D-region. Here, we can distinct between two typical cases: the refractive index is completely imaginary (EM wave attenuates), and completely real number (EM wave can propagate).
The first case holds for communication signal carrier frequencies below the electron plasma frequency which occurs in domains 96.63 kHz - 114.06 kHz, 142.19 kHz - 168.55 kHz and 209.21 kHz - 249.09 kHz at 70km, 75 km and 80 km, respectively, depending on temperature ({Fig. 5}). In the second case, when the value of the refractive index falls between 0 and 1, the electron plasma frequency is lower than the EM wave carrier frequency and an electromagnetic wave front experiences gradual refraction and bending. When the refractive index practically equals 1, the EM wave penetrates the considered ionospheric layer and reaches higher layers above.
In the transitional case $n=0$ when the electron plasma frequency is equal to the EM wave carrier frequency the signal blackout occurs. In Fig. 6, the color bar represents the real part value of $n$. The upper 3D plot shows the dependence of the real part of refractive index on the carrier frequency f and temperature T at three altitudes: 70 km, 75 km, and 80 km. The bottom picture shows the same dependence on T and f but in a more clear way as a contour  plot. Here we see that the altitude at which the real index of refraction $n$ takes a given value rises with the carrying signal frequency.
Also,  the refractive index decrease with rising temperature occurs at all three altitudes. These variations are more pronounced as altitude grows.

\section{Conclusions}

In this paper we present a new procedure for temperature calculations in the perturbed D-region, and the connection of the applied model with radio signal propagation characteristics and the D-region as the propagation medium.

The presented study is a theoretical procedure for determination of temperature time evolutions at the D-region heights from the effective recombination coefficient and expected temperature in absence of  perturbations which can be obtained from, for example, time and space dependent IRI model. This procedure is restricted to weak plasma changes in the considered time periods because of approximations of time constant ratios of recombination coefficients and densities of relevant ions.

\begin{wrapfigure}[18]{r}{0.5\textwidth}
\vspace{-20pt}
\begin{center}
\includegraphics[width=0.5\textwidth,height=0.5\textwidth]{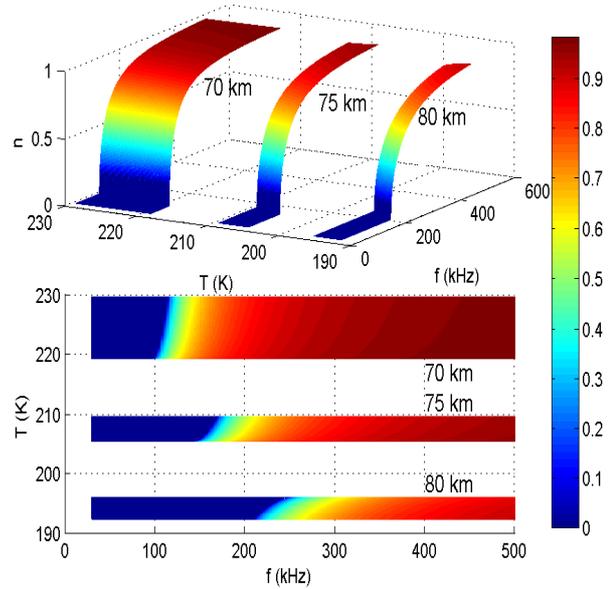}
\end{center}
\vspace{-27pt}
\caption{\small{Dependencies of refractive index from electron temperature and EM wave carrier
frequency at altitudes 70 km, 75 km and 80 km during relaxation.}}
\vspace{-10pt}
\label{Fig6_T_i_f}
\end{wrapfigure}

In this paper, we also give the procedure for calculations of the effective recombination coefficient from real time observations of the D-region by the VLF radio signal method. This monitoring technique allows determination of final data relevant to the considered time periods and locations. In this analysis, the calculations of the effective recombination coefficient are restricted to periods after large perturbations that induce changes in the D-region plasma parameters. Also, the given calculations require the absence of other sudden perturbations. Here, it is important to say that these two restrictions are related only to the presented calculations of effective recombination coefficient e.g. they are not required for further determination of temperature.

In addition to temperature calculations, we also give temperature connections with parameters important in telecommunications: the electron plasma density (representing medium properties important for radio signal propagation) and refractive index (representing signal propagation properties).


The obtained analytical equations are applied to a particular solar X-flare influence on the part of the D-region where the VLF signal emitted in Germany and received in Serbia propagate. The obtained final results show:

\begin{itemize}
\item Slow temperature relaxation during the 20 min period after important influence of the X-radiation on ionization processes in the D-region.
\item Decrease in temperature changes with  latitude.
\item More intensive dependency of electron plasma frequencies on temperature at higher altitudes.
\item Different changes in influences of the X-radiation on temperature and electron plasma frequencies with altitude: contrary to first one, changes of the second parameter increase with altitude.
\item The EM wave carrier frequency of a signal suffering a complete signal energy absorption (blackout) decreases during the relaxation period, e. g. it decreases with temperature decrease.
\item The changes in the refractive index of telecommunication signals are more visible at higher altitudes. Its values close to 1 occur from signal frequency that increase with altitude and temperature.
\end{itemize}

Finally, we want to point out that this study is expansion of investigations of solar activity induced temperature variation and investigation of telecommunication signal propagation under perturbed conditions at the D-region heights.

\section{Nomenclature}

$\Delta A _{\rm {rec}}$ \textendash Amplitude of received VLF radio signal, [dB]

\noindent $\Delta P _{\rm {rec}}$ \textendash Phase of received VLF radio signal, [$^{\rm 0}$]

\noindent $I$ \textendash X-radiation flux, [W/m$^{\rm 2}$]

\noindent $\alpha_{eff}$ \textendash Effective recombination coefficient, [m$^{\rm 3}$s$^{\rm {-1}}$]

\noindent $\alpha_{i}$ \textendash Recombination coefficient related to specific ion specie, [m$^{\rm 3}$s$^{\rm -1}$]

\noindent $N_e$ \textendash Electron plasma density, [m$^{\rm {-3}}$]

\noindent $N_i$ \textendash Ion plasma density, [m$^{\rm {-3}}$]

\noindent $C_i$ \textendash Coefficient related to ion specie, [\textendash]

\noindent $D_i$ \textendash Coefficient related to ion specie, [\textendash]

\noindent $T_e$ \textendash Electron temperature, [K]

\noindent $\alpha_{Claster}$ \textendash Recombination coefficient related to specific claster ion, [m$^{\rm 3}$s$^{\rm {-1}}$]

\noindent $r_{\alpha_{i}}^{Claster}$\textendash Specific and claster ions recombination coefficients ratio, [\textendash]

\noindent $r_{N_{i}}^{Claster}$ \textendash Specific and claster ions density ratio, [\textendash]

\noindent $\alpha_{eff}^{0}$ \textendash Effective recombination coefficient under unperturbed conditions, [m$^{{\rm 3}}$s$^{\rm {-1}}$]

\noindent $T_e^{0}$ \textendash Electron temperature under unperturbed conditions, [K]

\noindent ${\cal G}_0$ \textendash Electron gain rate, [m$^{\rm {-3}}$s$^{\rm {-1}}$]

\noindent $H'$ \textendash Signal reflection height, [km]

\noindent $\beta$ \textendash Sharpness, [km$^{ \rm {-1}}$]

\noindent $h$ \textendash Altitude, [km]

\noindent $t$ \textendash Time, [s]

\noindent $f_0$ \textendash Electron plasma frequency, [H$_{\rm z}$]

\noindent $f$ \textendash Communication radio signal frequency, [H$_{\rm z}$]

\noindent $n$ \textendash Refractive index, [\textendash]

\section*{Acknowledgment}

The authors would like to thank the Ministry of Education, Science and Technological Development of the Republic of Serbia for the support of this work within the projects III-44002, 176002, 176004 and TR-32030 and Ministry of Defence within the project VA-TT/OS5/2015. The authors are also grateful to the reviewers for helpful suggestions.

\bibliographystyle{unsrt}

\end{document}